\documentclass{aa}

\usepackage{graphicx}
\usepackage{txfonts}
\usepackage{lipsum}
\usepackage{subcaption}         % necessary for continued figures, example in section 3
\usepackage{lscape}             % to rotate a single page table, example in appendix.
\usepackage{placeins}           % useful with \FloatBarrier, to keep 
\usepackage{txfonts}
\usepackage{xcolor}
\usepackage{comment}

\titlerunning{MIDIS. Near-IR morphology at 3<z<5} 
\authorrunning{Costantin et al.}
 
\begin{document} 

   \title{MIDIS. Near-infrared rest-frame morphology
   of massive galaxies\\ at $3<z<5$ in the Hubble eXtreme Deep Field}

   \author{
        L.~Costantin\inst{1} 
        \and
        S.~Gillman\inst{2,3}
        \and
        L.~A.~Boogaard\inst{4}
        \and
        P.~G.~P\'erez-Gonz\'alez\inst{1}    
        \and
        E.~Iani\inst{5} 
        \and
        P.~Rinaldi\inst{5}
        \and        
        J.~Melinder\inst{6} 
	\and
        A.~Crespo G\'omez\inst{1}
        \and
        L.~Colina\inst{1}
        \and
        T.~R.~Greve\inst{2,3,7}
        \and
        G.~{\"O}stlin\inst{6}
        \and
        G.~Wright\inst{8}
        \and
        A.~Alonso-Herrero\inst{9}
        \and
        J.~\'Alvarez-M\'arquez\inst{1}
        \and
        M.~Annunziatella\inst{1} 
        \and
        A.~Bik\inst{6}
        \and
        K.~I.~Caputi\inst{2,5}
        \and
        D.~Dicken\inst{8}         
        \and
        A.~Eckart\inst{10}     
        \and
        J.~Hjorth\inst{11}
        \and
        O.~Ilbert\inst{12}
        \and 
        I.~Jermann\inst{2,3}
        \and
        A.~Labiano\inst{13}
        \and
        D.~Langeroodi\inst{11}
        \and
        F.~Pei{\ss}ker\inst{10}  
        \and
        J.~P.~Pye\inst{14}
        \and
        T.~V.~Tikkanen\inst{14}
        \and
        P.~P.~van der Werf\inst{15}      
        \and
        F.~Walter\inst{4}
        \and
        M.~Ward\inst{16}
        \and 
        M. G{\"u}del\inst{17,18}
        \and
        T.~K.~Henning\inst{4}
        }

   \institute{
        $^{1}$ Centro de Astrobiolog\'ia (CAB), CSIC-INTA, Ctra. de Ajalvir km 4, Torrej\'on de Ardoz, E-28850, Madrid, Spain\\
        $^{2}$ Cosmic Dawn Center (DAWN), Denmark\\
        $^{3}$ DTU-Space, Technical University of Denmark, Elektrovej 327, DK-2800 Kgs. Lyngby, Denmark\\
        $^{4}$ Max Planck Institute for Astronomy, K{\"o}nigstuhl 17, 69117 Heidelberg, Germany\\
        $^{5}$ Kapteyn Astronomical Institute, University of Groningen, P.O. Box 800, 9700AV Groningen, The Netherlands\\
        $^{6}$ Department of Astronomy, Stockholm University, Oscar Klein Centre, AlbaNova University Centre, 106 91 Stockholm, Sweden\\
        $^{7}$ Dept. of Physics and Astronomy, University College London, Gower Street, London WC1E 6BT, United Kingdom\\
        $^{8}$ UK Astronomy Technology Centre, Royal Observatory Edinburgh, Blackford Hill, Edinburgh EH9 3HJ, UK\\
        $^{9}$ Centro de Astrobiolog\'ia (CAB), CSIC-INTA, Camino Bajo del Castillo s/n, E-28692 Villanueva de la Ca\~nada, Madrid, Spain\\
        $^{10}$ I.Physikalisches Institut der Universit{\"a}t zu K{\"o}ln, Z{\"u}lpicher Str. 77, 50937 K{\"o}ln, Germany\\
        $^{11}$ DARK, Niels Bohr Institute, University of Copenhagen, Jagtvej 128, 2200 Copenhagen, Denmark\\
        $^{12}$ Aix Marseille Universit\'e, CNRS, LAM (Laboratoire d’Astrophysique de Marseille) UMR 7326, 13388, Marseille, France\\
        $^{13}$ Telespazio UK for the European Space Agency, ESAC, Camino Bajo del Castillo s/n, 28692 Villanueva de la Cañada, Spain\\
        $^{14}$ School of Physics \& Astronomy, Space Park Leicester, University of Leicester, 92 Corporation Road, Leicester LE4 5SP, UK\\
	$^{15}$ Leiden Observatory, Leiden University, P.O. Box 9513, 2300 RA Leiden, The Netherlands\\
	$^{16}$ Centre for Extragalactic Astronomy, Durham University, South Road, Durham DH1 3LE, UK\\
	$^{17}$ University of Vienna, Department of Astrophysics T{\"u}rkenschanzs- trasse 17, 1180 Vienna, Austria\\
	$^{18}$ Institute of Particle Physics and Astrophysics, ETH Zurich, Wolfgang-Pauli-Str 27, 8093 Zurich, Switzerland\\
        \email{lcostantin@cab.inta-csic.es}
             }

  \abstract
   {Thanks to decades of observations 
   using the Hubble Space Telescope (HST), the structure 
   of galaxies at redshift $z>2$ has been widely studied
   in the rest-frame ultraviolet regime, which traces recent star formation 
   from young stellar populations. But, we still have little information
   about the spatial distribution of the older, more evolved stellar populations,
   constrained by the rest-frame infrared portion of the galaxies' 
   spectral energy distribution.}
   {We present the morphological characterization of a sample 
   of 49 massive galaxies ($\log(M_{\star}/M_{\odot})>9$) at redshift $3<z<5$.
   These galaxies are observed as part of the Guaranteed Time Observations 
   program MIDIS with the MIRI instrument onboard JWST.
   The deep MIRI 5.6~$\mu$m imaging (28.64~mag $5\sigma$ depth) allows us to
   characterize the rest-frame near-infrared structure of galaxies
   beyond cosmic noon, at higher redshifts than possible with NIRCam, tracing their older 
   and dust-insensitive stellar populations.
   }
   {We derive the non-parametric morphology of galaxies,
   focusing on the Gini, $M_{20}$, concentration, asymmetry, and deviation statistics.
   Furthermore, we model the light distribution of galaxies with a single S\'ersic 
   component and derive their parametric morphology (i.e., effective radius and S\'ersic index).}
   {We find that at $z>3$ massive galaxies show a smooth distribution
   of their rest-infrared light, strongly supporting the 
   increasing number of regular disk galaxies 
   already in place at early epochs. These results
   are further reinforced by the analysis of JWST/NIRCam data at 4.4~$\mu$m.
   On the contrary, the ultraviolet structure obtained from HST/WFC3 and JWST/NIRCam
   observations at $\sim1.5~\mu$m is generally more irregular,
   catching the most recent episodes of star formation.
   Importantly, we find a segregation of morphologies across cosmic time, where galaxies 
   at redshift $z>3.75$ show later-type morphologies compared to $z\sim3$ galaxies.
   These findings suggest a transition phase in galaxy assembly and central mass build-up, 
   which is already taking place at $z\sim3-4$.}
    {The combined analysis of NIRCam and MIRI imaging datasets
   allows us to prove that the rest-frame near-infrared morphology of
   massive galaxies at cosmic noon is typical of 
   compact disk galaxies with a smooth mass distribution.}


   \maketitle
%
%-------------------------------------------------------------------

\section{Introduction}

Galaxy morphology is a key proxy of galaxy diversity,
since it provides a first glimpse of
the physical processes involved in galaxy evolution.
Indeed, the structural evolution of galaxies across cosmic time
is shown to be strongly related to stellar mass and star 
formation history, merger history, and environment
\citep{Visvanathan.N:1977, Tully.B:1982, Kennicutt.R:1998, Kauffmann.G:2003, 
Baldry.I:2004, DeLucia.G:2007, Lotz.J:2008, Blanton.M:2009, Kormendy.J:2010}.

In the last two decades, our knowledge of galaxy structure 
beyond the local Universe was based on studies making use
of the Hubble Space Telescope (HST), which has given us access 
to the rest-frame ultraviolet to optical morphology
of galaxies up to redshift $z\sim2-3$.
At these wavelengths, galaxies at $z>1$ 
appear more irregular in their light distribution
than local galaxies \citep{Conselice.C:2000, Conselice.C:2008},
and such peculiar systems dominate 
the galaxy population beyond $z \sim 2.5$
\citep[e.g.,][]{Buitrago.F:2013, HuertasCompany.M:2015}.

Based on almost 30 years of HST observational campaigns, 
multiple results pointed to the conclusion that the Hubble sequence 
is established around $z\gtrsim1$
\citep{Brinchmann.J:1998, Faber.S:2007, Bruce.V:2012, Barro.G:2013, Mortlock.A:2013, HuertasCompany.M:2016}.
But, properly quantifying the morphological transformation of galaxies
using HST datasets suffers some limitations. A combination of
insufficient spatial resolution and limited red wavelength coverage
makes it difficult to properly resolve the first complex structures assembling
in the first $2-3$~Gyr after the Big Bang (i.e., $z>2-3$).
Thus, this observational limitation leaves open the question whether
the Hubble sequence was already in place at earlier cosmic times
then previously thought.
Indeed, spectroscopic observations already suggest
an epoch of early disk assembly \citep[e.g., ][]{Wisnioski.E:2015, 
Simons.R:2017, Rizzo.F:2020},
with the first complex structures starting to build up
at and beyond cosmic noon \citep[e.g.,][]{Tacchella.S:2015,
Costantin.L:2021, Costantin.L:2022, 
Yunpeng.J:2024A, Jegatheesan.K:2024}.

Now, for the first time, the James Webb Space 
Telescope \citep[JWST;][]{Gardner.J:2023}
opens the possibility to characterize 
the detailed structure of the bulk
of the stellar population of the highest-redshift galaxies
($z>3$) with an unprecedented level of detail 
\citep[e.g.,][]{Ferreira.L:2022, Ferreira.L:2023,
Kartaltepe.J:2023, Costantin.L:2023b, Treu.T:2023, 
Jacobs.C:2023, HuertasCompany.M:2025}.
The prevailing conclusion of these initial studies 
is that the fraction of galaxies with disk-like morphologies 
is higher than that inferred with HST,
although their exact nature still needs to be investigated
\citep[e.g.,][]{VegaFerrero.J:2024, Pandya.V:2024}.

In this context, the Mid-Infrared Instrument 
\citep[MIRI;][]{Rieke.G:2015, Bouchet.P:2015, Wright.G:2023, Dicken.D:2024} 
onboard JWST,
probing the observed near-to-mid infrared regime (4.9 to 27.9~$\mu$m), 
provides a huge jump in sensitivity compared to previous observatories
at these wavelengths \citep[$\sim$10 times deeper 
than IRAC/Spitzer;][]{Fazio.G:2004}.
MIRI allows targeting the rest-frame near-infrared
morphology of galaxies at $z>3$ (up to $z\sim5$), resolving structures
at a few kpc scale ($FWHM \sim 0.2$~arcsec at 5.6~$\mu$m, 
corresponding to $\sim1.6$~kpc at $z=3$ and $\sim1.3$~kpc at $z=5$).
Furthermore, MIRI sensitivity could also allow us to
probe the near-infrared
morphology of galaxies at $5<z<10$ (i.e., F770W and F1000W bands;
$FWHM \sim 0.3$~arcsec at 10~$\mu$m, 
corresponding to $\sim1.4$~kpc at $z=10$).

At high redshift, the MIRI Deep Imaging Survey 
\citep[MIDIS;][]{Ostlin.G:2024}
is the best available dataset to address key open questions 
about the clumpy/irregular distribution of the bulk 
of the stellar mass,
as traced by the older and almost dust-insensitive stellar populations
\citep[see e.g.,][Gillman et al.~in prep.]{Boogaard.L:2024}.
It consists of $\sim41.34$~hours of net exposure time
for deep imaging of the Hubble eXtreme Deep Field 
\citep[XDF;][]{Illingworth.GD:2013}
at 5.6~$\mu$m, with parallel observations
of the surrounding area \citep[see e.g.,][]{PerezGonzalez.P:2023}
using the Near Infrared Camera (NIRCam)
and the Near Infrared Imager and Slitless Spectrograph (NIRISS).
Furthermore, 8.5~hours of net exposure time
were dedicated to deep imaging in the F1000W band
\citep[][]{PerezGonzalez.P:2024b, Iani.E:2024b}.

In this work, we study the near-infrared
rest-frame morphologies of massive galaxies at redshift $z>3$
in the XDF using MIDIS observations.
We derive their non-parametric and parametric morphology,
characterizing the structure of their old stellar population
and looking at the morphological transformation of the bulk 
of their stellar mass across cosmic time.
The paper is organized as follows. In Sect.~\ref{sec:section2}
we describe the selection of the sample of galaxies. 
In Sect.~\ref{sec:section3} we derive the 
non-parametric and parametric 
morphology, discussing the implications in the context of 
galaxy evolution. Finally, in Sect.~\ref{sec:section4}
we summarize our results and provide our conclusions.

Throughout this work we assume a \citet{Planck2018} cosmology with
$H_0 = 67.4$~km~s$^{-1}$~Mpc$^{-1}$, $\Omega_m = 0.315$, 
and $\Omega_{\Lambda} = 0.685$. 
We quote magnitudes in the AB system \citep{Oke.J:1983} and
all errors are reported as the 16th–84th percentile interval.

\begin{figure*}[t!]
\centering
\includegraphics[width=0.80\textwidth]{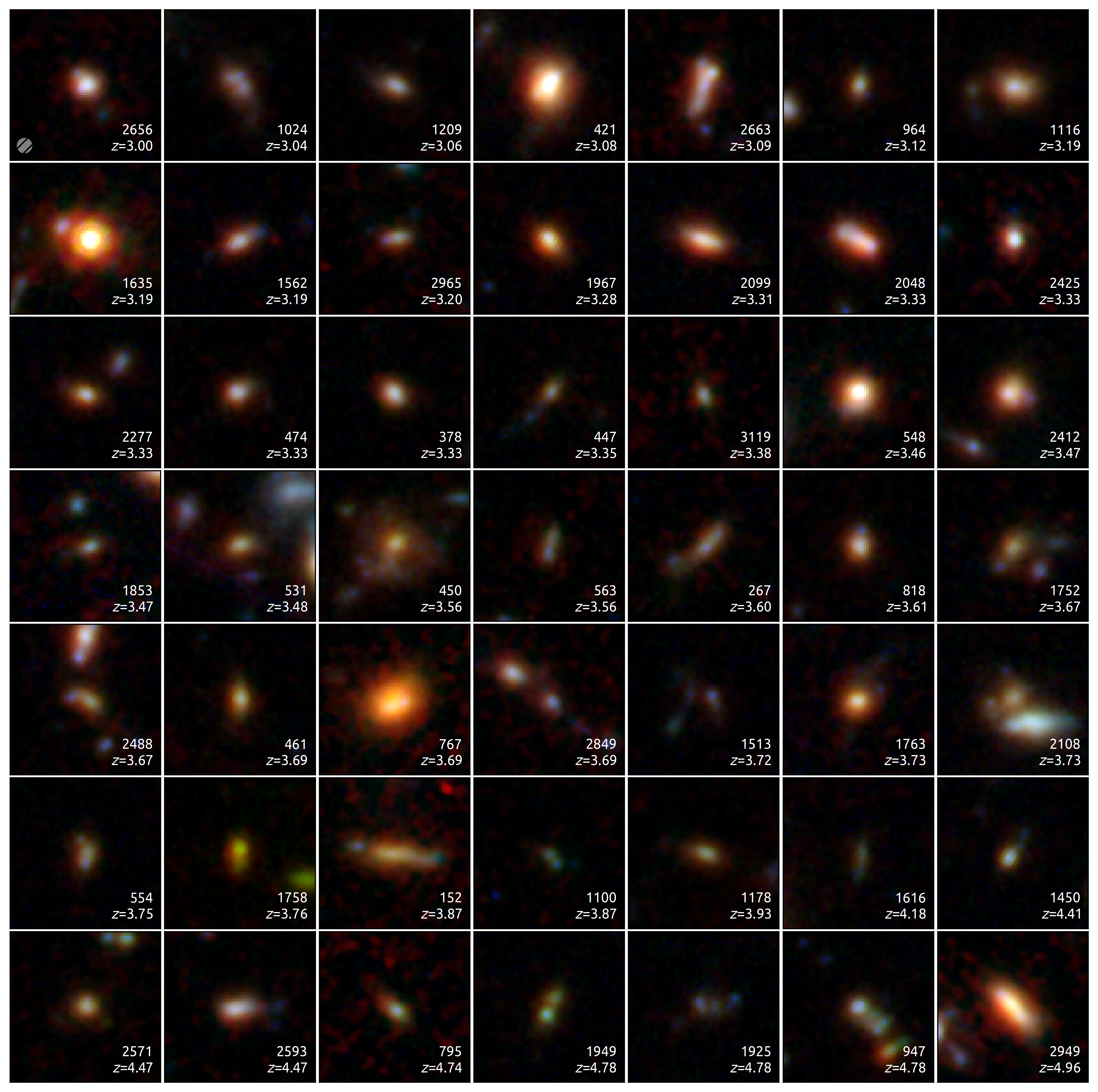}
\caption{MIRI/F560W, NIRCam/F356W, and NIRCam/F150W RGB images
of the 49 galaxies studied in this work,
ordered by increasing redshift.
The cutouts are $2 \times 2$~arcsec$^2$.
We report the angular resolution as the FWHM of the MIRI PSF.
\label{fig:figure1}}
\end{figure*}

%_____________________________________________________________
%                     Long table using the longtab environment
%-------------------------------------------------------------
\longtab{
{\small
\begin{longtable}{c c c c c c c c c c c}
\caption{Sample galaxies and morphological statistics at 5.6~$\mu$m, ordered by increasing redshift.}\\
\label{tab:table1}
ID & RA & DEC & $z$ & Stellar Mass & Gini & $M_{20}$ & $C$  & $A$  & $R_{\rm e}$ & $n$ \\
& [degree] & [degree] & & [$\log(M_{\star}/M_{\odot}$)] & & & & & [kpc] & \\
(1) & (2) & (3) & (4) & (5) & (6) & (7) & (8) & (9) & (10) & (11) \\
\hline
\endhead
\hline
\endfoot
2656   &    $53.19078$   &    $-27.77812$    &    $3.00^{\dagger}$           &    $9.19$     &    $0.444_{-0.007}^{+0.006}$    &    $-1.60_{-0.02}^{+0.02}$    &    $2.37_{-0.02}^{+0.02}$    &    $0.11_{-0.01}^{+0.01}$         &    $0.48_{-0.03}^{+0.05}$    &    $5_{-2}^{+3}$    \\
1024   &    $53.16501$   &    $-27.78786$    &    $3.04_{-0.03}^{+0.03}$  &    $9.14$     &    $0.47_{-0.01}^{+0.01}$          &    $-1.67_{-0.03}^{+0.03}$    &    $2.51_{-0.03}^{+0.03}$    &    $0.18_{-0.01}^{+0.02}$         &    $2.6_{-0.3}^{+0.3}$          &    $2.7_{-0.4}^{+0.4}$    \\
1209   &    $53.16739$   &    $-27.78532$    &    $3.06^{\dagger}$            &    $9.02$     &    $0.47_{-0.02}^{+0.02}$          &    $-1.71_{-0.03}^{+0.03}$    &    $2.55_{-0.03}^{+0.03}$    &    $0.13_{-0.01}^{+0.01}$         &    $0.96_{-0.09}^{+0.12}$    &    $4_{-1}^{+2}$    \\
421     &    $53.15260$   &    $-27.79392$    &    $3.08^{\dagger}$            &    $9.93$     &    $0.503_{-0.006}^{+0.005}$    &    $-1.67_{-0.01}^{+0.01}$    &    $2.55_{-0.01}^{+0.01}$    &    $0.087_{-0.007}^{+0.007}$    &    $0.92_{-0.01}^{+0.01}$    &    $1.70_{-0.08}^{+0.08}$    \\
2663   &    $53.16996$   &    $-27.76842$    &    $3.09^{\dagger}$            &    $9.76$     &    $0.46_{-0.03}^{+0.03}$          &    $-1.1_{-0.1}^{+0.1}$          &    $2.4_{-0.1}^{+0.1}$          &    $0.26_{-0.02}^{+0.03}$          &    $1.77_{-0.05}^{+0.06}$    &    $0.61^{\star}$    \\
964     &    $53.16709$   &    $-27.79027$    &    $3.12_{-0.06}^{+0.09}$  &    $9.01$     &    $0.47_{-0.01}^{+0.01}$          &    $-1.62_{-0.03}^{+0.03}$    &    $2.45_{-0.03}^{+0.0}$      &    $0.10_{-0.02}^{+0.02}$          &    $0.50_{-0.04}^{+0.05}$    &    $1.3_{-0.4}^{+0.5}$    \\
1116   &    $53.16968$   &    $-27.78816$    &    $3.19^{\dagger}$            &    $9.36$      &    $0.427_{-0.007}^{+0.007}$    &    $-1.57_{-0.02}^{+0.02}$    &    $2.31_{-0.02}^{+0.02}$    &    $0.10_{-0.01}^{+0.01}$          &    $1.11_{-0.03}^{+0.03}$    &    $1.11_{-0.13}^{+0.11}$    \\
1635   &    $53.17851$   &    $-27.78411$    &    $3.19^{\dagger}$            &    $10.11$    &    $0.555_{-0.004}^{+0.005}$    &    $-1.79_{-0.01}^{+0.01}$    &    $2.86_{-0.01}^{+0.01}$    &    $0.100_{-0.005}^{+0.006}$    &    $-$                                   &    $-$    \\
1562   &    $53.17668$   &    $-27.78388$    &    $3.19^{\dagger}$            &    $9.33$     &    $0.47_{-0.01}^{+0.01}$          &    $-1.57_{-0.02}^{+0.02}$    &    $2.39_{-0.02}^{+0.02}$     &    $0.15_{-0.02}^{+0.01}$         &    $0.91_{-0.06}^{+0.06}$    &    $1.1_{-0.3}^{+0.4}$    \\
2965   &    $53.17175$   &    $-27.76673$    &    $3.20_{-0.07}^{+0.03}$  &    $9.09$     &    $0.374_{-0.02}^{+0.02}$         &    $-1.59_{-0.04}^{+0.04}$    &    $2.17_{-0.04}^{+0.04}$    &    $0.14_{-0.03}^{+0.03}$          &    $0.9_{-0.2}^{+0.4}$          &    $5_{-2}^{+3}$    \\
1967   &    $53.17461$   &    $-27.77794$    &    $3.28^{\dagger}$            &    $9.65$     &    $0.531_{-0.006}^{+0.006}$     &    $-1.71_{-0.02}^{+0.02}$    &    $2.55_{-0.02}^{+0.02}$    &    $0.089_{-0.009}^{+0.008}$    &    $0.45_{-0.01}^{+0.01}$    &    $1.5_{-0.2}^{+0.2}$    \\
2099   &    $53.16181$   &    $-27.77072$    &    $3.31_{-0.07}^{+0.02}$  &    $9.71$     &    $0.488_{-0.005}^{+0.007}$     &    $-1.67_{-0.01}^{+0.01}$    &    $2.52_{-0.01}^{+0.01}$    &    $0.107_{-0.008}^{+0.008}$    &    $1.00_{-0.02}^{+0.02}$    &    $1.2_{-0.1}^{+0.1}$    \\
2048   &    $53.16286$   &    $-27.77170$    &    $3.33^{\dagger}$            &    $9.93$     &    $0.468_{-0.006}^{+0.007}$     &    $-1.38_{-0.01}^{+0.01}$    &    $2.36_{-0.01}^{+0.01}$    &    $0.260_{-0.008}^{+0.008}$    &    $1.10_{-0.02}^{+0.01}$    &    $0.80_{-0.07}^{+0.08}$    \\
2425   &    $53.19576$   &    $-27.78279$    &    $3.33^{\dagger}$            &    $9.96$     &    $0.421_{-0.008}^{+0.010}$     &    $-1.62_{-0.02}^{+0.02}$    &    $2.26_{-0.02}^{+0.02}$    &    $0.11_{-0.02}^{+0.01}$          &    $0.4_{-0.01}^{+0.02}$      &    $0.8_{-0.2}^{+0.3}$    \\
2277   &    $53.16978$   &    $-27.77258$    &    $3.33^{\dagger}$            &    $9.24$     &    $0.466_{-0.007}^{+0.009}$     &    $-1.62_{-0.03}^{+0.03}$    &    $2.52_{-0.03}^{+0.02}$    &    $0.11_{-0.01}^{+0.01}$          &    $0.56_{-0.05}^{+0.06}$    &    $5_{-1}^{+2}$    \\
474     &    $53.15311$   &    $-27.79246$    &    $3.33_{-0.05}^{+0.02}$   &    $9.25$     &    $0.475_{-0.009}^{+0.008}$     &    $-1.42_{-0.02}^{+0.02}$    &    $2.40_{-0.02}^{+0.02}$    &    $0.17_{-0.02}^{+0.02}$          &    $0.44_{-0.03}^{+0.04}$    &    $2.0_{-0.5}^{+0.6}$    \\
378     &    $53.14606$   &    $-27.79167$    &    $3.33^{\dagger}$            &    $9.38$     &    $0.444_{-0.006}^{+0.007}$     &    $-1.53_{-0.02}^{+0.02}$    &    $2.31_{-0.02}^{+0.02}$    &    $0.10_{-0.01}^{+0.01}$          &    $0.50_{-0.02}^{+0.03}$    &    $1.1_{-0.3}^{+0.3}$    \\
447     &    $53.14686$   &    $-27.79035$    &    $3.35_{-0.01}^{+0.03}$   &    $9.00$     &    $0.44_{-0.03}^{+0.03}$           &    $-1.52_{-0.06}^{+0.07}$    &    $2.42_{-0.06}^{+0.07}$    &    $0.14_{-0.03}^{+0.02}$           &    $0.8_{-0.2}^{+0.2}$          &    $4_{-2}^{+2}$    \\
3119   &    $53.19433$   &    $-27.77584$    &    $3.38_{-0.23}^{+0.02}$   &    $9.17$     &    $0.43_{-0.04}^{+0.04}$           &    $-1.40_{-0.08}^{+0.08}$    &    $2.14_{-0.08}^{+0.08}$    &    $0.17_{-0.04}^{+0.04}$           &    $0.58_{-0.10}^{+0.15}$    &    $4_{-2}^{+3}$    \\
548     &    $53.16792$   &    $-27.79803$    &    $3.46^{\dagger}$            &    $9.85$     &    $0.499_{-0.005}^{+0.006}$     &    $-1.63_{-0.01}^{+0.02}$    &    $2.54_{-0.01}^{+0.02}$    &    $0.104_{-0.008}^{+0.009}$     &    $0.48_{-0.01}^{+0.01}$    &    $2.2_{-0.3}^{+0.4}$    \\
2412   &    $53.17413$   &    $-27.77305$    &    $3.47_{-0.02}^{+0.04}$   &    $9.69$     &    $0.45_{-0.01}^{+0.20}$           &    $-1.62_{-0.01}^{+0.01}$    &    $2.37_{-0.01}^{+0.01}$    &    $0.098_{-0.008}^{+0.008}$    &    $0.66_{-0.02}^{+0.02}$     &    $1.5_{-0.2}^{+0.2}$    \\
1853   &    $53.15641$   &    $-27.77074$    &    $3.47^{\dagger}$            &    $9.19$     &    $0.45_{-0.04}^{+0.04}$           &    $-1.62_{-0.08}^{+0.08}$    &    $2.14_{-0.08}^{+0.08}$    &    $0.14_{-0.03}^{+0.04}$          &    $0.87_{-0.24}^{+0.33}$     &    $3_{-1}^{+2}$    \\
531     &    $53.15700$   &    $-27.79445$    &    $3.48_{-0.02}^{+0.01}$  &    $10.51$   &    $0.48_{-0.05}^{+0.05}$           &    $-1.68_{-0.08}^{+0.08}$    &    $2.59_{-0.08}^{+0.08}$    &    $0.09_{-0.03}^{+0.03}$          &    $6_{-2}^{+2}$                    &    $8^{\star}$    \\
450     &    $53.14922$   &    $-27.79156$    &    $3.56^{\dagger}$            &    $9.89$     &    $0.38_{-0.02}^{+0.02}$           &    $-1.64_{-0.04}^{+0.03}$    &    $2.50_{-0.04}^{+0.03}$    &    $0.20_{-0.02}^{+0.02}$          &    $2.6_{-0.2}^{+0.2}$           &    $2.1_{-0.2}^{+0.2}$    \\
563     &    $53.14677$   &    $-27.78772$    &    $3.56^{\dagger}$            &    $9.06$     &    $0.34_{-0.03}^{+0.03}$           &    $-1.36_{-0.06}^{+0.07}$    &    $1.94_{-0.06}^{+0.07}$    &    $0.16_{-0.04}^{+0.03}$          &    $1.1_{-0.2}^{+0.2}$           &    $1.0_{-0.3}^{+0.8}$    \\
267     &    $53.16088$   &    $-27.80120$    &    $3.60^{\dagger}$            &    $9.50$     &    $0.46_{-0.06}^{+0.04}$           &    $-1.59_{-0.06}^{+0.06}$    &    $2.29_{-0.06}^{+0.06}$    &    $0.13_{-0.03}^{+0.02}$          &    $1.9_{-0.1}^{+0.1}$           &    $0.7_{-0.1}^{+0.2}$    \\
818     &    $53.15157$   &    $-27.78541$    &    $3.61^{\dagger}$            &    $9.46$     &    $0.499_{-0.009}^{+0.008}$     &    $-1.64_{-0.02}^{+0.02}$    &    $2.51_{-0.02}^{+0.02}$    &    $0.09_{-0.01}^{+0.01}$          &    $0.60_{-0.05}^{+0.06}$     &    $4_{-1}^{+1}$    \\
1752   &    $53.17936$   &    $-27.78299$    &    $3.67^{\dagger}$            &    $9.23$     &    $0.40_{-0.01}^{+0.02}$           &    $-1.53_{-0.03}^{+0.03}$    &    $2.31_{-0.03}^{+0.03}$    &    $0.22_{-0.02}^{+0.02}$          &    $1.20_{-0.09}^{+0.10}$     &    $1.2_{-0.3}^{+0.4}$    \\
2488   &    $53.18854$   &    $-27.77871$    &    $3.67^{\dagger}$            &    $9.15$     &    $0.36_{-0.01}^{+0.01}$           &    $-1.51_{-0.03}^{+0.02}$    &    $2.08_{-0.03}^{+0.02}$    &    $0.14_{-0.02}^{+0.02}$          &    $1.11_{-0.07}^{+0.08}$     &    $1.0_{-0.3}^{+0.4}$    \\
461     &    $53.16566$   &    $-27.79862$    &    $3.69_{-0.08}^{+0.05}$  &    $9.17$     &    $0.50_{-0.02}^{+0.01}$           &    $-1.63_{-0.02}^{+0.03}$    &    $2.47_{-0.02}^{+0.03}$    &    $0.15_{-0.01}^{+0.02}$           &    $0.68_{-0.08}^{+0.06}$    &    $3.2_{-0.9}^{+1.1}$    \\
767     &    $53.14815$   &    $-27.78449$    &    $3.69_{-0.03}^{+0.02}$  &    $10.60$   &    $0.483_{-0.005}^{+0.005}$     &    $-1.65_{-0.01}^{+0.02}$    &    $2.50_{-0.01}^{+0.02}$    &    $0.098_{-0.008}^{+0.009}$     &    $1.12_{-0.03}^{+0.03}$    &    $2.1_{-0.2}^{+0.2}$    \\
2849   &    $53.18754$   &    $-27.77496$    &    $3.69_{-0.19}^{+0.03}$  &    $9.13$     &    $0.44_{-0.07}^{+0.15}$           &    $-1.4_{-0.5}^{+0.7}$          &    $2.2_{-0.5}^{+0.7}$          &    $0.2_{-0.3}^{+0.3}$                 &    $2_{-1}^{+2}$                   &    $7^{\star}$    \\
1513   &    $53.15798$   &    $-27.77585$    &    $3.72^{\dagger}$            &    $9.02$    &    $0.51_{-0.04}^{+0.03}$            &    $-1.43_{-0.08}^{+0.07}$    &    $2.48_{-0.08}^{+0.07}$    &    $0.25_{-0.04}^{+0.05}$           &    $1.2_{-0.1}^{+0.1}$          &    $1.1_{-0.3}^{+0.5}$    \\
1763   &    $53.15651$   &    $-27.77226$    &    $3.73^{\dagger}$            &    $9.50$    &    $0.458_{-0.007}^{+0.008}$      &    $-1.66_{-0.02}^{+0.02}$    &    $2.52_{-0.02}^{+0.02}$    &    $0.13_{-0.01}^{+0.01}$           &    $0.56_{-0.06}^{+0.09}$    &    $8^{\star}$    \\
2108   &    $53.16254$   &    $-27.77106$    &    $3.73^{\dagger}$            &    $9.49$    &    $0.49_{-0.02}^{+0.15}$            &    $-1.3_{-0.1}^{+0.1}$          &    $2.1_{-0.1}^{+0.1}$          &    $0.04_{-0.02}^{+0.02}$           &    $1.40_{-0.07}^{+0.07}$    &    $1.1_{-0.2}^{+0.2}$    \\
554     &    $53.16594$   &    $-27.79699$    &    $3.75_{-0.07}^{+0.05}$  &    $9.19$    &    $0.41_{-0.01}^{+0.01}$            &    $-1.62_{-0.03}^{+0.02}$    &    $2.22_{-0.03}^{+0.02}$    &    $0.17_{-0.02}^{+0.02}$           &    $0.95_{-0.07}^{+0.07}$    &    $0.69^{\star}$    \\
1758   &    $53.18674$   &    $-27.78634$    &    $3.76_{-0.02}^{+0.02}$  &    $9.04$    &    $0.41_{-0.01}^{+0.02}$            &    $-1.54_{-0.03}^{+0.03}$    &    $2.33_{-0.03}^{+0.03}$    &    $0.16_{-0.02}^{+0.02}$           &    $0.65_{-0.08}^{+0.06}$    &    $1.3_{-0.4}^{+0.5}$    \\
152     &    $53.15125$   &    $-27.79827$    &    $3.87_{-0.03}^{+0.04}$  &    $9.93$    &    $0.39_{-0.01}^{+0.01}$            &    $-1.36_{-0.06}^{+0.07}$    &    $2.59_{-0.06}^{+0.07}$    &    $0.18_{-0.02}^{+0.02}$           &    $3.1_{-0.3}^{+0.3}$          &    $1.3_{-0.3}^{+0.4}$    \\
1100   &    $53.15155$   &    $-27.77993$    &    $3.87_{-0.03}^{+0.04}$  &    $9.02$    &    $0.37_{-0.09}^{+0.11}$             &    $-1.3_{-0.2}^{+0.2}$          &    $1.8_{-0.2}^{+0.2}$          &    $0.12_{-0.06}^{+0.07}$          &    $0.6_{-0.1}^{+0.2}$           &    $2_{-1}^{+3}$    \\
1178   &    $53.15546$   &    $-27.78030$    &    $3.93_{-0.06}^{+0.06}$  &    $9.30$    &    $0.40_{-0.01}^{+0.02}$             &    $-1.61_{-0.03}^{+0.03}$    &    $2.33_{-0.03}^{+0.03}$    &    $0.10_{-0.01}^{+0.02}$          &    $5_{-2}^{+3}$                   &    $8^{\star}$    \\
1616   &    $53.17592$   &    $-27.78279$    &    $4.18^{\dagger}$            &    $9.04$    &    $0.37_{-0.02}^{+0.03}$            &    $-1.42_{-0.06}^{+0.05}$     &    $2.11_{-0.06}^{+0.05}$    &    $0.17_{-0.03}^{+0.03}$          &    $6_{-3}^{+6}$                   &    $8^{\star}$   \\
1450   &    $53.17333$   &    $-27.78389$    &    $4.41^{\dagger}$            &    $9.14$    &    $0.51_{-0.05}^{+0.03}$            &    $-1.38_{-0.06}^{+0.06}$     &    $2.41_{-0.06}^{+0.06}$    &    $0.13_{-0.02}^{+0.02}$         &    $0.42_{-0.01}^{+0.02}$    &    $0.67^{\star}$    \\
2571   &    $53.18086$   &    $-27.77420$    &    $4.47^{\dagger}$            &    $9.48$    &    $0.34_{-0.01}^{+0.01}$            &    $-1.49_{-0.03}^{+0.03}$     &    $2.07_{-0.03}^{+0.03}$    &    $0.12_{-0.02}^{+0.02}$         &    $0.46_{-0.04}^{+0.06}$    &    $4_{-2}^{+2}$    \\
2593   &    $53.18854$   &    $-27.77762$    &    $4.47^{\dagger}$            &    $9.14$    &    $0.37_{-0.02}^{+0.02}$            &    $-1.45_{-0.03}^{+0.02}$     &    $2.19_{-0.03}^{+0.02}$    &    $0.14_{-0.02}^{+0.02}$         &    $0.97_{-0.08}^{+0.07}$    &    $1.1_{-0.3}^{+0.5}$    \\
795     &    $53.17215$   &    $-27.79517$    &    $4.74_{-0.04}^{+0.04}$  &    $9.39$    &    $0.31_{-0.02}^{+0.03}$            &    $-1.42_{-0.07}^{+0.08}$    &    $2.05_{-0.07}^{+0.08}$    &    $0.15_{-0.03}^{+0.04}$           &    $0.96_{-0.24}^{+0.28}$    &    $3_{-1}^{+3}$    \\
1949   &    $53.16718$   &    $-27.77462$    &    $4.78^{\dagger}$            &    $9.14$    &    $0.385_{-0.009}^{+0.009}$      &    $-1.28_{-0.03}^{+0.03}$    &    $2.12_{-0.03}^{+0.03}$    &    $0.14_{-0.02}^{+0.02}$           &    $1.44_{-0.05}^{+0.05}$    &    $0.54^{\star}$    \\
1925   &    $53.16261$   &    $-27.77292$    &    $4.78^{\dagger}$            &    $9.06$    &    $0.47_{-0.05}^{+0.05}$            &    $-1.52_{-0.09}^{+0.08}$    &    $2.22_{-0.09}^{+0.08}$    &    $0.23_{-0.03}^{+0.03}$           &    $1.3_{-0.1}^{+0.1}$          &    $0.8_{-0.2}^{+0.3}$    \\
947     &    $53.15817$   &    $-27.78648$    &    $4.78^{\dagger}$            &    $9.36$    &    $0.43_{-0.01}^{+0.01}$            &    $-0.67_{-0.02}^{+0.03}$    &    $1.78_{-0.02}^{+0.03}$    &    $0.30_{-0.01}^{+0.02}$           &    $0.55_{-0.05}^{+0.07}$    &    $2.3_{-0.8}^{+1.4}$    \\
2949   &    $53.19873$   &    $-27.77972$    &    $4.96_{-0.02}^{+0.02}$  &    $10.80$   &    $0.419_{-0.007}^{+0.009}$     &    $-1.59_{-0.02}^{+0.02}$    &    $2.35_{-0.02}^{+0.02}$    &    $0.13_{-0.01}^{+0.01}$           &    $1.35_{-0.03}^{+0.02}$    &    $0.8_{-0.1}^{+0.1}$    \\
\end{longtable}
\tablefoot{Columns: (1) Galaxy ID. 
(2) Right ascension.
(3) Declination. 
(4) Photometric redshift (spectroscopic redshifts are marked with $^{\dagger}$). 
(5) Stellar mass \citep[typical uncertainties $\sim0.2$~dex; see e.g.,][]{Mobasher.B:2015}.
(6)-(9) Gini, $M_{20}$, $C$, and $A$ statistics.
(10) Parametric effective radius.
(11) Parametric S\'ersic index.
Galaxies marked by the $^{\star}$ symbol
have unreliable size or S\'ersic index, 
on the edge of the parameter space.}
}}

%--------------------------------------------------------------------
\section{Data and Sample \label{sec:section2}}

We select galaxies from the MIRI $5.6~\mu$m imaging 
of the JWST Guaranteed Time Observations (GTO) program MIDIS (PID: 1283),
which represents the deepest image of the Universe at these wavelengths 
\citep[28.64~mag $5\sigma$ depth, calculated within circular 
apertures with a diameter of 0.45~arcsec and corrected for 
drizzling correlation;][]{Rinaldi.P:2023, Rinaldi.P:2024, Iani.E:2024, Boogaard.L:2024}.
The data were calibrated using the procedure described in 
\citet{Ostlin.G:2024}, using a modified version 
of the official JWST pipeline version 1.12.3
(\texttt{pmap} 1137) and applying a background 
homogenization algorithm (including 1/f-noise removal) 
before obtaining the final mosaic
drizzled at a pixel scale of 0.06~arcsec
\citep[see also][for more details]{PerezGonzalez.P:2024a, PerezGonzalez.P:2024b}.
Due to the complex observational strategy of MIDIS,
and the lack of bright and isolated stars in the field,
we use a varying Point Spread Function (PSF) model
built using empirical PSFs at different positions on the MIRI detector
according to the MIDIS observational strategy \citep[see][]{Boogaard.L:2024},
oversampling models from \citet{Libralato.M:2024}.
As complementary datasets, we make use of imaging data
from CANDELS \citep{Grogin.N:2011, Koekemoer.A:2011} 
and JADES \citep{Rieke.M:2023, Eisenstein.D:2023},
covering observed wavelengths from $\sim 0.4-1.6$ to $1.1-4.4$~$\mu$m, respectively.

We detect and model galaxies in the F560W MIRI band,
and then forced photometry in HST, NIRCam, and MIRI bands is performed with
\texttt{THE FARMER} \citep{Weaver.J:2022},
allowing the flux to vary, whilst keeping the structural parameters fixed.
Photometric redshifts and stellar masses 
are derived from the multi-wavelength fluxes 
with \texttt{EAZY-PY} \citep{Brammer.G:2008}, employing
thirteen templates from the Flexible Stellar Populations Synthesis code 
\citep[FSPS;][]{Conroy.C:2010} as described in \citet{Kokorev.V:2022}.
The details about the source detection and photometric catalog 
are extensively described in Gillman et al.~\emph{in prep}.

Galaxies are initially selected for having redshift $3<z<5$
and stellar masses $\log(M_{\star}/M_{\odot})>9$.
The total number of such galaxies in the MIDIS field is 67.
Visually inspecting the images of the initial sample, 
we discard ten faint galaxies (S/N<5),
four galaxies that are at the edge of the MIRI pointing,
and four galaxies that are not resolved or 
are extremely contaminated by foreground sources.
The main properties of the sample galaxies are enumerated in Table~\ref{tab:table1}.

With this selection, we are probing for the first time
the rest-frame regime $\gtrsim0.9$~$\mu$m
for 49 galaxies up to $z=5$, which is unique of the MIDIS dataset
and where there is essentially no variation 
of galaxies' morphology \citep[e.g.,][]{Martorano.M:2023, Ren.J:2024}.
It is finally worth noticing that galaxy 767 
is presented in the ALMA selected sample analyzed in 
\citet{Boogaard.L:2024}, while galaxies 1635 and 2663 
are in the sample of X-ray Active Galactic Nuclei detailed in
\citet{Gillman.S:2025}.

%-----------------------------------------------------------------
\section{Method and Results \label{sec:section3}}

In this work, we present the non-parametric and parametric
morphology for a sample of 49 massive galaxies 
at $z>3$ using MIRI imaging at 5.6~$\mu$m.
Given the redshift range of our sample ($3 < z < 5$),
we are probing the rest-frame near-infrared structure of galaxies
($\lambda_{\rm rest} \sim 0.9-1.4$~$\mu$m),
which best traces more evolved stellar populations
that only MIRI can probe up to this high redshift.
To complement our analysis, we compare the rest-frame 
near-infrared structure of these galaxies with their 
rest-frame ultraviolet spatial distribution using
HST/F160W imaging from CANDELS.
Apart from the different stellar populations probed
by these two bands, our choice is also 
justified by the similar width of the PSF of these two datasets,
minimizing effects related to spatial resolutions
(see also Appendix~\ref{sec:appendixA}).

\subsection{Non-parametric morphology \label{sec:section3.1}}

We measure the structure of the sample galaxies observed in the MIRI/F560W filter.
Firstly, we create segmentation maps of each galaxy with \texttt{SEP}
\citep{Barbary.K:2018}, a Python library implementing 
\texttt{SExtractor} \citep{Bertin.E:1996}.
Then, we derive non-parametric morphological diagnostics 
using \texttt{statmorph} \citep{RodriguezGomez.V:2019}.
Following \citet{Crespo.A:2024}, we quantify the 
uncertainties associated with each parameter
by performing 500 Monte Carlo realizations for each galaxy.
We perturb each image pixel-by-pixel with Gaussian noise 
($\sigma = $ root mean square of the background level)
and create new segmentation maps at each iteration.
The 16th-84th percentile interval of each parameter is used to
quantify their uncertainty.

\begin{figure}[t!]
\centering
\includegraphics[width=0.47\textwidth]{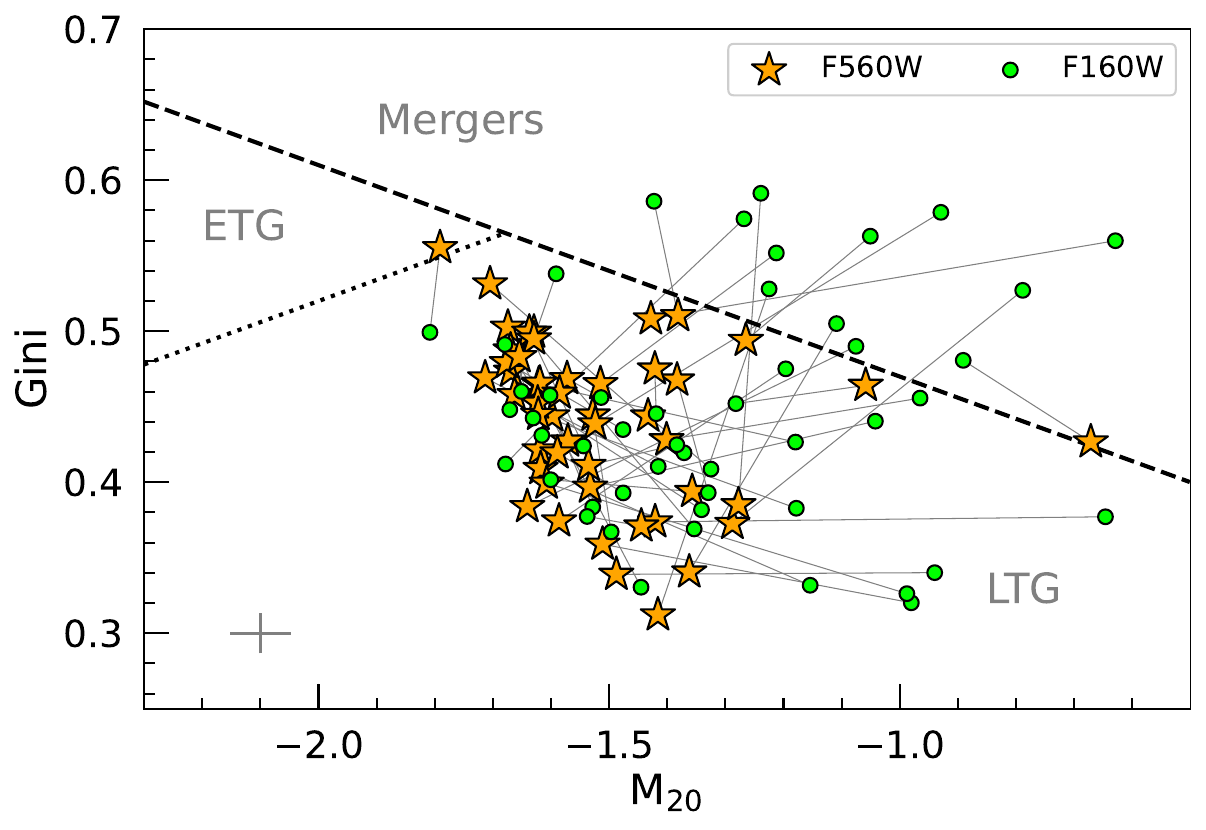}
\caption{Gini-$M_{20}$ diagram. 
Orange stars and green dots correspond to non-parametric morphology
measured on MIRI F560W and WFC3 F160W images, respectively. 
For each galaxy, gray lines link the observed morphology at different wavelengths.
ETGs, LTGs, and mergers are separated according to Eqs.~(4) in \citet{Lotz.J:2008}.
\label{fig:figure2}}
\end{figure}

In the following, we focus on the Gini and $M_{20}$ 
statistics (see Table~\ref{tab:table1} 
and Figs.~\ref{fig:figure2}, \ref{fig:figure3}, and \ref{fig:figure5}).
The Gini coefficient quantifies the light
distribution of the galaxy, ranging from $G=0$
for homogeneous brightness distribution
to $G=1$ when the entire flux is concentrated 
in a single pixel \citep{Abraham.R:2003, Lotz.J:2004}.
The $M_{20}$ statistic is the
normalized second-order moment of the brightest 20\% 
of the galaxy’s flux and it provides valuable information
about the spatial distribution of any sub-structure,
such as bright nuclei, bars, or spiral arms \citep{Lotz.J:2004}.
Thus, the Gini-$M_{20}$ diagram has been largely employed 
to separate early, late-type, and merging 
galaxies from low to high redshift 
\citep{Lotz.J:2004, Lotz.J:2008, Rose.C:2023, Crespo.A:2024}.

Furthermore, to better characterize the sample galaxies,
using \texttt{statmorph} we derive the 
concentration– asymmetry–smoothness
\citep[CAS;][]{Conselice.C:2003} 
and multimode–intensity–deviation statistics
\citep[MID;][]{Freeman.P:2013}, which allow
us better discriminate the true nature of 
galaxies in the Gini-$M_{20}$ diagram.
In particular, we complement the information
inferred from the Gini-$M_{20}$ diagram
with that provided by the concentration, asymmetry, and deviation
statistics (see Table~\ref{tab:table1} and Figs.~\ref{fig:figure4}-\ref{fig:figure5}).
The concentration index is defined as the 
ratio of the radii that enclose 80\% and 20\% 
of the light of the galaxy \citep{Conselice.C:2003, Lotz.J:2004}.
As a consequence, elliptical galaxies are the most concentrated systems, 
and the concentration decreases for later Hubble types \citep{Bershady.M:2000}. 
The asymmetry coefficient quantifies 
the degree to which the light distribution of the galaxy is rotationally symmetric.
Generally, the asymmetry is more sensitive to merger signatures 
than concentration, with irregular galaxies more asymmetric than disky or spheroidal galaxies.
It is worth mentioning that we report the asymmetry coefficients without accounting
for the asymmetry of the background, which dominates our measurements and leads
to artificially low or even negative asymmetry values \citep[see e.g.,][]{Bignone.L:2020}.
The deviation index \citep{Freeman.P:2013} quantifies
the variation of irregular or peculiar morphologies from elliptical symmetry.
In this case, it is expected that elliptical or regular disk galaxies
with no substructures (e.g., clumps) show values of deviation clustering near zero.

It is worth noting that while both Gini and $M_{20}$
usually correlate with concentration,
they differ in important respects.
On one side, the Gini coefficient is independent of the 
large-scale spatial distribution of light, and high values 
of Gini may also arise if the bright structures 
are not located in the center of a galaxy (unlike $C$).
On the other side, $M_{20}$ (scaling as $r^2$)
is more heavily weighted than $C$
by the spatial distribution of bright regions.
Moreover, $M_{20}$ is more sensitive than $C$
to merger signatures, since it is not measured within
circular or elliptical apertures with a fixed center
\citep[see][for further details]{Lotz.J:2004}.

\subsubsection{Gini-$M_{20}$ \label{sec:section3.1.1}}

In Fig.~\ref{fig:figure2}, we show the distribution
of rest-frame near-infrared morphologies in the (Gini, $M_{20}$) 
diagram (orange stars: see also Fig.~\ref{fig:figure_A2}). 
The first result is that all galaxies in our sample occupy the region of 
late-type systems, even if some 
(i.e., 947, 1450, 1513, 1635, 2108, and 2663) are
quite close to the edge of mergers or early-type objects.
Looking at their visual morphology (Fig.~\ref{fig:figure1}),
we can confirm that all these galaxies but 1450 have very close companions
that contaminate their surface-brightness profiles (see also Fig.~\ref{fig:figure5}).

As a comparison, we derive the rest-frame ultraviolet morphologies
of the sample galaxies using HST/WFC3 imaging from CANDELS 
(Gini$_{\rm UV}$ and $M_{20, \rm UV}$; green points).
Furthermore, we analyze NIRCam images from JADES
at similar wavelengths (F150W) but different spatial resolution (see Fig.~\ref{fig:figure_A1}).
Already at $z>3$, galaxies appear more irregular at
shorter wavelengths, pointing to the presence of an underlying already mature 
population with a smooth stellar mass distribution
already in place when the Universe was $\lesssim2$~Gyr old.
These results are confirmed by the analysis of NIRCam datasets
at higher spatial resolution, as shown in Appendix~\ref{sec:appendixA}.

We find that in the rest-frame near-infrared regime (F560W) galaxies have
Gini~$= 0.45_{-0.07}^{+0.05}$ and $M_{20} = -1.57_{-0.09}^{+0.20}$,
while in the ultraviolet regime (F160W) they have
Gini$_{\rm UV} = 0.44_{-0.06}^{+0.09}$ and $M_{20, \rm UV} = -1.3_{-0.3}^{+0.4}$.
For individual galaxies, we derive median
$\Delta$Gini $= 0.01_{-0.12}^{+0.06}$ and $\Delta$M$_{20} = -0.2_{-0.3}^{+0.2}$.
Thus, galaxies move from later-type morphologies 
to earlier-type ones if observed at shorter or longer wavelengths, respectively.
This trend is consistent with previous results based on
ultraviolet and optical differences of typical
star-forming galaxies at $z\sim3-4$ reported in
\citet{Conselice.C:2008} and \citet{Wuyts.S:2012}, 
and could be explained as evidence for disk assembly 
through the inward migration of clumps and gas accretion.

\begin{figure}[t!]
\centering
\includegraphics[width=0.467\textwidth]{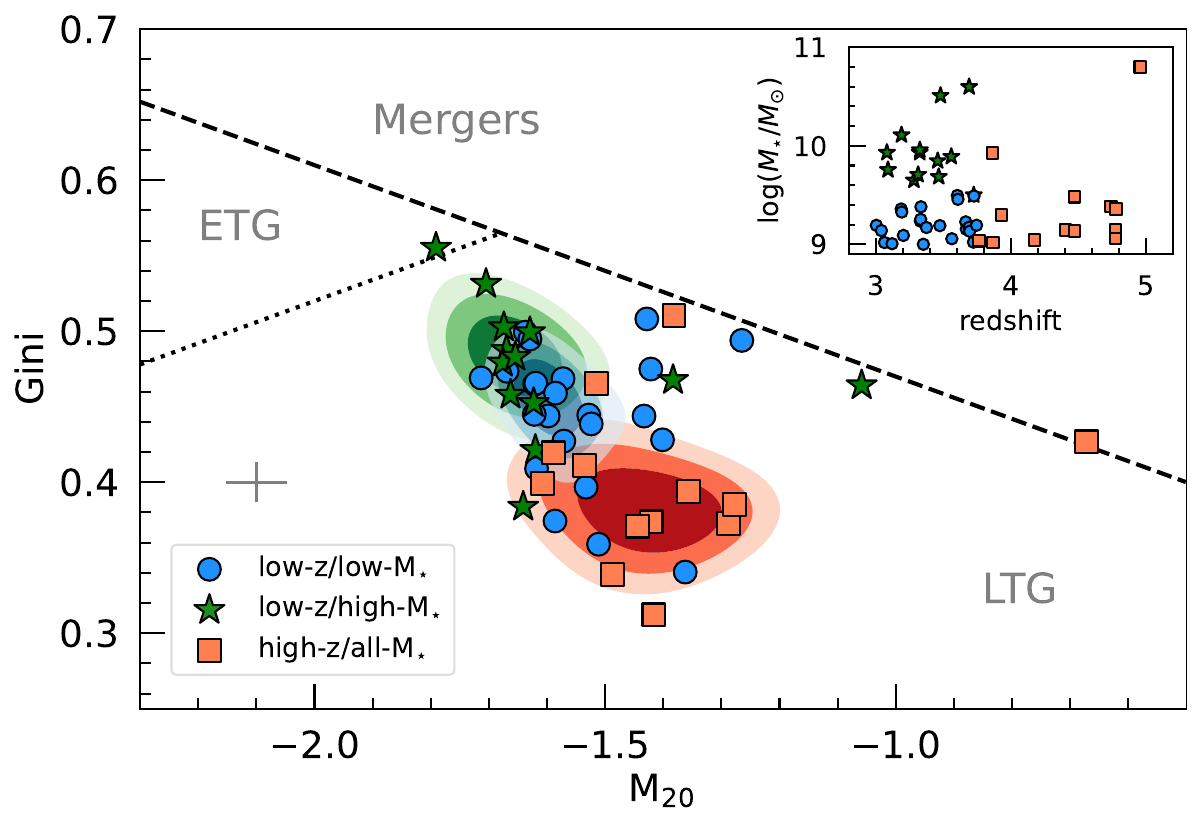}
\caption{Gini-$M_{20}$ diagram based on MIRI F560W morphology, 
separating galaxies at $3 < z < 3.75$ with $\log(M_{\star}/M_{\odot})<9.5$ 
(\texttt{low-z/low-$M_{\star}$}; blue dots and shaded region),
galaxies at $3 < z < 3.75$ with $\log(M_{\star}/M_{\odot})>9.5$ 
(\texttt{low-z/high-$M_{\star}$}; green stars and shaded region),
and galaxies at $3.75 < z < 5$ (\texttt{high-z/all-$M_{\star}$}; red squares and shaded region).
ETGs, LTGs, and mergers are separated according to Eqs.~(4) in \citet{Lotz.J:2008}.
\label{fig:figure3}}
\end{figure}

We explore if there is any trend with redshift or stellar mass,
separating galaxies into two redshift bins of equivalent lookback time of 
$\sim 1$~Gyr. Furthermore, we separate galaxies in the low-redshift bin according to
their stellar mass. Thus, the three classes are defined as it follows. 
\texttt{low-z/low-$M_{\star}$} has $3 < z < 3.75$ and $\log(M_{\star}/M_{\odot})<9.5$,
\texttt{low-z/high-$M_{\star}$} has $3 < z < 3.75$ and $\log(M_{\star}/M_{\odot})>9.5$,
while \texttt{high-z/all-$M_{\star}$} has $3.75 < z < 5$ (all masses).
In Fig.~\ref{fig:figure3}, we show the Gini, $M_{20}$ diagram
for the three different classes.
\texttt{low-z/low-$M_{\star}$} galaxies show median Gini~$= 0.45_{-0.07}^{+0.04}$ 
and $M_{20} = -1.57_{-0.06}^{+0.15}$, \texttt{low-z/high-$M_{\star}$} galaxies have median
Gini~$= 0.48_{-0.03}^{+0.03}$
and $M_{20} = -1.65_{-0.02}^{+0.06}$,
while \texttt{high-z/all-$M_{\star}$} galaxies have median Gini~$= 0.39_{-0.03}^{+0.04}$
and $M_{20} = -1.4_{-0.1}^{+0.1}$.
We see that galaxies in the higher redshift bin appear more
irregular than galaxies at lower redshift,
suggesting a segregation of morphologies 
from irregular to smoother light (and stellar mass) profiles.
A similar transition is also seen accounting for the different stellar mass of 
galaxies at $3 < z < 3.75$, even if the trend is milder.

\subsubsection{Concentration, asymmetry, and deviation \label{sec:deviation}}

To better quantify the structure of the sample galaxies,
we derive their concentration, asymmetry,
and deviation statistics.

\begin{figure}[t!]
\centering
\includegraphics[width=0.49\textwidth]{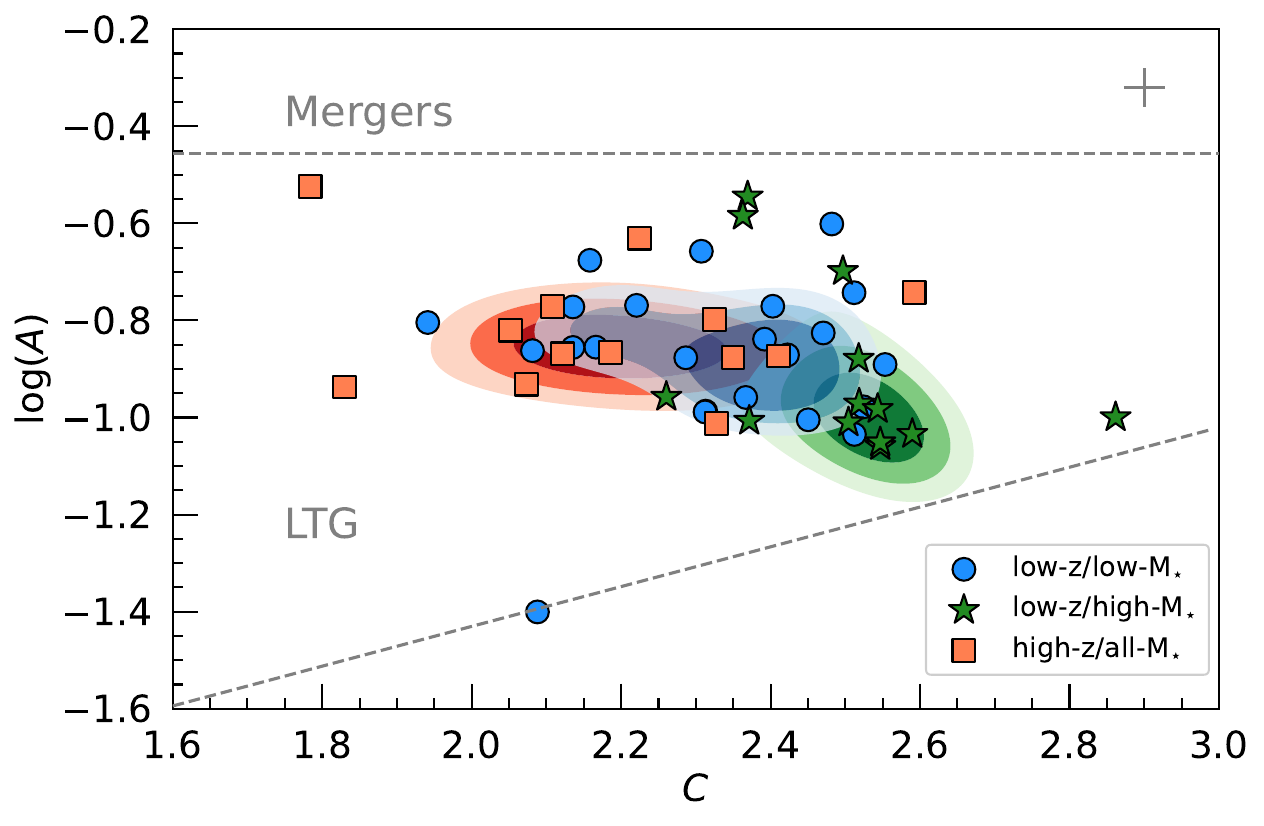}
\caption{$C$-$A$ diagram based on MIRI F560W morphology, 
separating galaxies at $3 < z < 3.75$ with $\log(M_{\star}/M_{\odot})<9.5$ 
(\texttt{low-z/low-$M_{\star}$}; blue dots and shaded region),
galaxies at $3 < z < 3.75$ with $\log(M_{\star}/M_{\odot})>9.5$ 
(\texttt{low-z/high-$M_{\star}$}; green stars and shaded region),
and galaxies at $3.75 < z < 5$ (\texttt{high-z/all-$M_{\star}$}; red squares and shaded region).
Mergers and late-type galaxies are separated according to \citet{Conselice.C:2003},
the boundaries between late-type galaxies 
and intermediate galaxies are defined as in \citet{Bershady.M:2000}.
\label{fig:figure4}}
\end{figure}

\begin{figure*}[t!]
\centering
\includegraphics[width=0.95\textwidth]{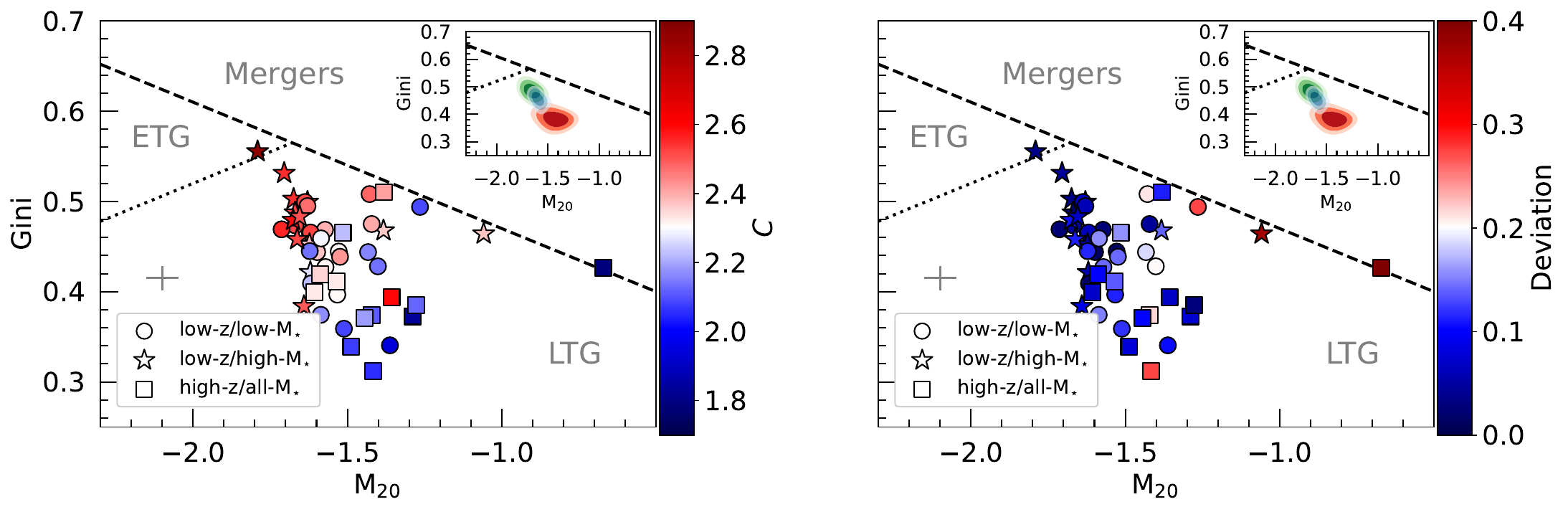}
\caption{Gini-$M_{20}$ diagram, color-coded according 
to concentration (left panel) and deviation (right panel).
Galaxies are divided in \texttt{low-z/low-$M_{\star}$} (dots),
\texttt{low-z/high-$M_{\star}$} (stars), and \texttt{high-z/all-$M_{\star}$} (squares).
The inset plot shows the density distribution 
as reported in Fig.~\ref{fig:figure3}.
\label{fig:figure5}}
\end{figure*}

In Fig.~\ref{fig:figure4}, we show the three classes of galaxies
in the $C$-$A$ diagnostic \citep{Bershady.M:2000, Conselice.C:2003}. 
We find that all the aforementioned results are confirmed, 
with \texttt{low-z/low-$M_{\star}$} galaxies having 
median $C = 2.3_{-0.2}^{+0.2}$ 
and $A = 0.14_{-0.03}^{+0.04}$, \texttt{low-z/high-$M_{\star}$} galaxies showing
median $C = 2.52_{-0.15}^{+0.03}$ 
and $A = 0.10_{-0.01}^{+0.10}$, and \texttt{high-z/all-$M_{\star}$} presenting
median $C = 2.2_{-0.2}^{+0.2}$ 
and $A = 0.14_{-0.02}^{+0.05}$.

In the Gini, $M_{20}$ diagram (Fig.~\ref{fig:figure5}),
the concentration statistic reflects the trend 
previously highlighted, with lower redshift galaxies (all masses)
being more concentrated (median $C = 2.4_{-0.2}^{+0.1}$) than higher
redshift ones (median $C = 2.2_{-0.2}^{+0.2}$). As expected, the galaxy's 
concentration increases moving from LTGs to ETGs along 
empirical trend Gini~$\propto -M_{20}$, which is also used to 
separate normal from merging systems \citep{Lotz.J:2008}. 
On the other hand, the deviation statistic allows us
to identify more irregular morphologies,
such as 795, 947, 1513, 1616, 2108, 2663, and 3119 (Fig.~\ref{fig:figure1}).

\subsection{Parametric morphology}

We performed the two-dimensional photometric decomposition 
of the MIRI images, modeling the surface brightness 
of each galaxy with a \citet{Sersic.J:1968} law.
We make use of \texttt{anduryl} (Marrero de la Rosa et al.~\emph{in prep.})
a computational tool designed for 2D photometric fitting of galaxies,
leveraging Bayesian inference as its foundational framework.
The software accommodates fits based on either a S\'ersic model or a 
combination of S\'ersic and exponential models.
At the heart of \texttt{anduryl} lies a robust 
parameter space exploration, 
executed through Nested Sampling, a technique for approximating the 
posterior probability integral implemented 
through Nestle\footnote{\url{http://kylebarbary.com/nestle/index.html}}. 
This method enables the inference of the 
posterior probability distribution, 
thereby facilitating the extraction of marginal posterior distributions 
for each parameter. From these distributions, the mean values emerge as 
the most plausible estimates for each parameter.

\begin{figure*}[t!]
\centering
\includegraphics[width=0.95\textwidth]{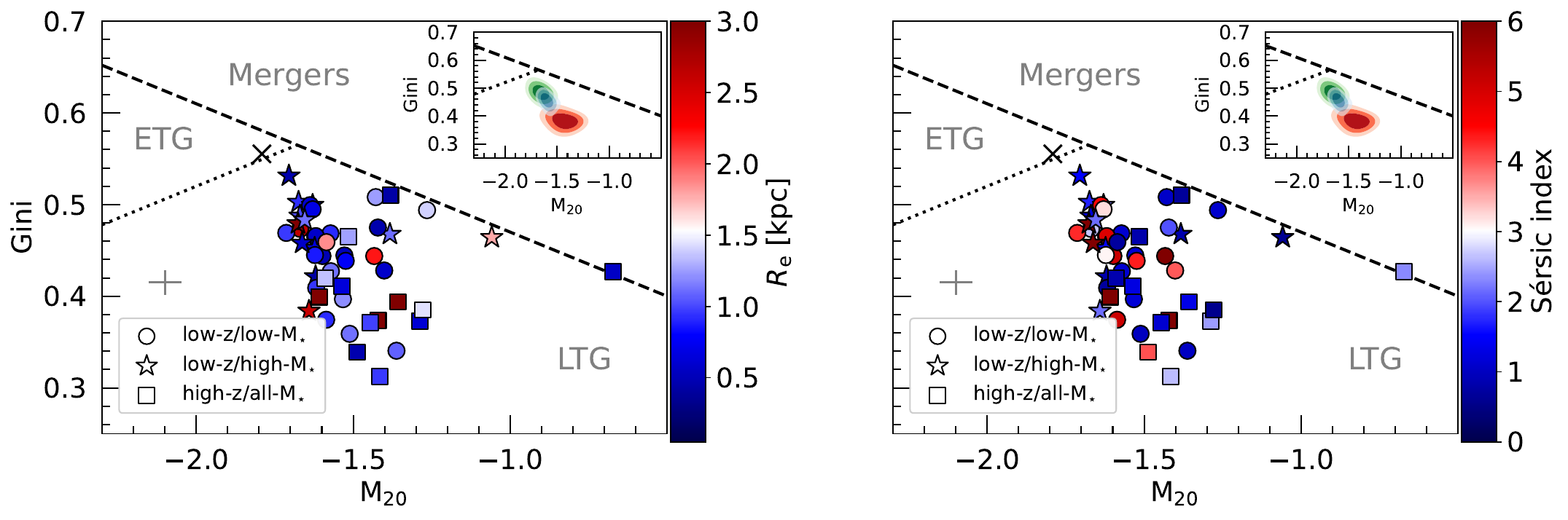}
\caption{As Fig.~\ref{fig:figure5}, but for 
the effective radius (left panel) and S\'ersic index (right panel).
Galaxy ID~1635 is marked with an "X",
since it has no parametric model.
\label{fig:figure6}}
\end{figure*}

Looking at the size (half-light radius $R_{\rm e}$)
and S\'ersic index of the sample galaxies,
we find no clear correlation with the Gini-M$_{20}$ statistics (Fig.~\ref{fig:figure6}).
In detail, \texttt{low-z/low-$M_{\star}$} galaxies have median $R_{\rm e} = 0.9_{-0.4}^{+0.4}$~kpc 
and S\'ersic index $n = 2.0_{-0.9}^{+2.7}$, 
\texttt{low-z/high-$M_{\star}$} galaxies show median $R_{\rm e} = 0.9_{-0.5}^{+0.9}$~kpc 
and S\'ersic index $n = 1.5_{-0.8}^{+1.2}$,
while \texttt{high-z/all-$M_{\star}$} galaxies show median $R_{\rm e} = 1.3_{-0.7}^{+2.5}$~kpc 
and S\'ersic index $n = 1.3_{-0.6}^{+3.1}$.
Overall, the median values for the entire sample are 
$R_{\rm e} = 1.0_{-0.5}^{+1.1}$~kpc 
and $n = 1.5_{-0.7}^{+3.2}$, which could be 
considered typical of late-type morphologies,
but with a wide scatter to high values.

%-----------------------------------------------------------------

\section{Summary and Conclusions \label{sec:section4}}

In this work, we present the first detailed rest-frame near-infrared 
morphological study of a sample of 49 galaxies at $3 < z < 5$ 
with $\log(M_{\star}/M_{\odot}) > 9$, 
observed as part of the MIRI Deep Imaging Survey 
in the XDF at 5.6~micron (F560W).

We employ non-parametric morphological diagnostics
to classify the sample galaxies, mainly focusing 
on the Gini and $M_{20}$, as well as concentration, asymmetry,
deviation, and parametric Sérsic modelling of the galaxies.

From this study, we can draw three main conclusions.
First, massive galaxies at $z>3$ show more regular
structures in the restframe near-infrared regime (MIRI F560W)
compared to a more peculiar morphology
at shorter wavelengths. We measure the morphology 
of galaxies using the F150W and F160W bands 
(NIRCam/JEST and WFC3/HST, respectively), probing the
restframe ultraviolet regime of these galaxies,
finding that their structure is more irregular.
Second, the visual, non-parametric (Gini-$M_{20}$, concentration, asymmetry,
deviation), and parametric (S\'ersic and $R_{\rm e}$)
analysis points to a compact population of disk-like 
galaxies, with a mostly regular mass distribution.
The analysis of higher resolution images 
at 4.4~$\mu$m (see Appendix~\ref{sec:appendixA}) 
reinforces our conclusion that massive galaxies at $3<z<5$ 
show disk-like morphologies with a smooth mass distribution
and a quite compact structure.
Third, we separate galaxies into three classes.
\texttt{low-z/low-$M_{\star}$} galaxies have 
$3 < z < 3.75$ and $\log(M_{\star}/M_{\odot})<9.5$, 
\texttt{low-z/high-$M_{\star}$} galaxies 
have $3 < z < 3.75$ and $\log(M_{\star}/M_{\odot})>9.5$,
while \texttt{high-z/all-$M_{\star}$} have $3.75 < z < 5$ (all masses).
We find a segregation 
of galaxy morphologies across cosmic time 
in the Gini-$M_{20}$ and $C$-$A$ diagrams, from later to earlier types.
This could be interpreted as
a transition phase in galaxy assembly,
where the complex structures (i.e., bulges, bars) start to assemble first.

Building on the results of this work, 
deep MIRI campaigns in the near future 
will open the possibility to expand the morphological characterization 
of galaxies in the early Universe 
to a larger sample and possibly to a higher redshift 
(e.g., up to $z=10$ using MIRI F770W and F1000W).

\begin{acknowledgements}

We would like to thank I.~Smail for all the comments that improved the content of the manuscript
and C.~J.~Conselice for the useful discussion.

The project that gave rise to these results received the support of a fellowship from the 
``la Caixa” Foundation (ID 100010434). The fellowship code is LCF/BQ/PR24/12050015.
LC and PGPG acknowledge support from grant PID2022-139567NB-I00 funded by Spanish Ministerio de Ciencia 
e Innovaci\'on MCIN/AEI/10.13039/501100011033, FEDER \emph{Una manera de hacer Europa}.
LC, JA-M, AC-G, and LC acknowledge support by grant PIB2021-127718NB-100 from the Spanish Ministry of Science and Innovation/State Agency of Research MCIN/AEI/10.13039/501100011033 and by “ERDF A way of making Europe”.
SG acknowledges financial support from the Villum Young Investigator grant 37440 and 13160 and the Cosmic Dawn Center (DAWN), funded by the Danish National Research Foundation under grant DNRF140.
LAB acknowledges support by ERC AdG grant 740246 (Cosmic-Gas).
EI acknowledges funding from the Netherlands Research School for Astronomy (NOVA).
JM and AB acknowledge support from the Swedish National Space Administration (SNSA).
AAH acknowledges support from grant PID2021-124665NB-I00 funded by MCIN/AEI/10.13039/501100011033 and by ERDF A way of making Europe”. AE and 
KIC acknowledges funding from the Netherlands Research School for Astronomy (NOVA) and from the Dutch Research Council (NWO) through the award of the Vici Grant VI.C.212.036.
IJ acknowledge support from the Carlsberg Foundation (grant no CF20-0534) and the Cosmic Dawn Center is funded by the Danish National Research Foundation under grant No.~140.
FP acknowledges support through the German Space Agency DLR 50OS1501 and DLR 50OS2001 from 2015 to 2023.
JPP and TVT acknowledge financial support from the UK Science and Technology Facilities Council, and the UK Space Agency.
This work was supported by research grants (VIL16599, VIL54489) from VILLUM FONDEN.

The work presented is the effort of the entire MIRI team and the
enthusiasm within the MIRI partnership is a significant factor in
its success. MIRI draws on the scientific and technical expertise of
the following organisations: Ames Research Center, USA; Airbus
Defence and Space, UK; CEA-Irfu, Saclay, France; Centre Spatial de
Li\'ege, Belgium; Consejo Superior de Investigaciones Científicas,
Spain; Carl Zeiss Optronics, Germany; Chalmers University of
Technology, Sweden; Danish Space Research Institute, Denmark; Dublin
Institute for Advanced Studies, Ireland; European Space Agency,
Netherlands; ETCA, Belgium; ETH Zurich, Switzerland; Goddard Space
Flight Center, USA; Institute d'Astrophysique Spatiale, France;
Instituto Nacional de T\'ecnica Aeroespacial, Spain; Institute for
Astronomy, Edinburgh, UK; Jet Propulsion Laboratory, USA;
Laboratoire d'Astrophysique de Marseille (LAM), France; Leiden
University, Netherlands; Lockheed Advanced Technology Center (USA);
NOVA Opt-IR group at Dwingeloo, Netherlands; Northrop Grumman, USA;
Max-Planck Institut für Astronomie (MPIA), Heidelberg, Germany;
Laboratoire d’Etudes Spatiales et d'Instrumentation en Astrophysique
(LESIA), France; Paul Scherrer Institut, Switzerland; Raytheon
Vision Systems, USA; RUAG Aerospace, Switzerland; Rutherford
Appleton Laboratory (RAL Space), UK; Space Telescope Science
Institute, USA; Toegepast- Natuurwetenschappelijk Onderzoek
(TNO-TPD), Netherlands; UK Astronomy Technology Centre, UK;
University College London, UK; University of Amsterdam, Netherlands;
University of Arizona, USA; University of Cardiff, UK; University of
Cologne, Germany; University of Ghent; University of Groningen,
Netherlands; University of Leicester, UK; University of Leuven,
Belgium; University of Stockholm, Sweden; Utah State University,
USA. A portion of this work was carried out at the Jet Propulsion
Laboratory, California Institute of Technology, under a contract
with the National Aeronautics and Space Administration. We would
like to thank the following National and International Funding
Agencies for their support of the MIRI development: NASA; ESA;
Belgian Science Policy Office; Centre Nationale D'Etudes Spatiales
(CNES); Danish National Space Centre; Deutsches Zentrum fur Luft-und
Raumfahrt (DLR); Enterprise Ireland; Ministerio De Econom\'ia y
Competitividad; Netherlands Research School for Astronomy (NOVA);
Netherlands Organisation for Scientific Research (NWO); Science and
Technology Facilities Council; Swiss Space Office; Swedish National
Space Board; UK Space Agency.
For the purpose of open access, the author has applied a Creative Commons Attribution 
(CC BY) licence to the Author Accepted Manuscript version arising from this submission.
    
\end{acknowledgements}

\begin{appendix}
\section{Gini-$M_{20}$ at 1.5 and 4.4~$\mu$m \label{sec:appendixA}}

In this Appendix, we complement the results presented in Sect.~\ref{sec:section3.1.1}
deriving the Gini and $M_{20}$ diagnostics in the UV and near-IR regime through 
the analysis of NIRCam images in the F150W and F444W bands.

\begin{figure}[t!]
\centering
\includegraphics[width=0.47\textwidth]{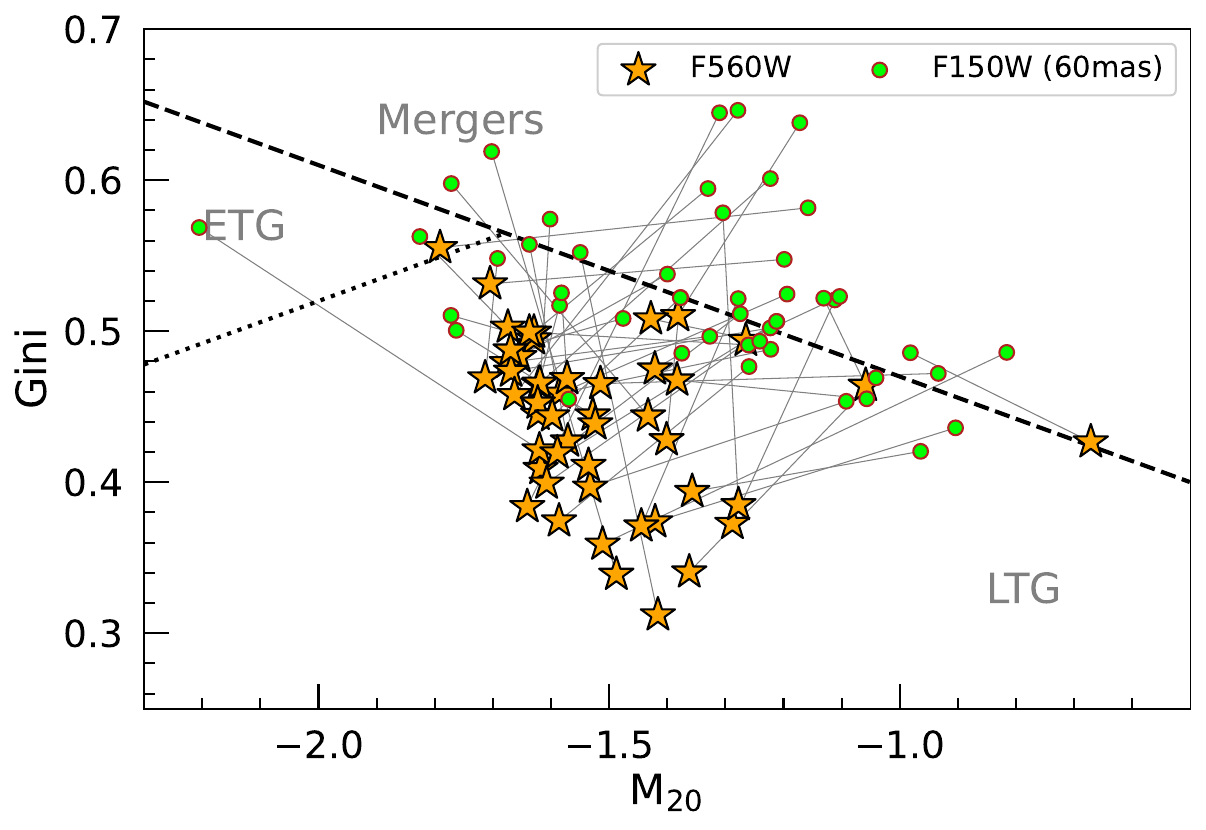}
\caption{As Fig.~\ref{fig:figure2}, but for MIRI F560W (orange stars) and NIRCam F150W images (green dots). 
\label{fig:figure_A1}}
\end{figure}

\begin{figure}[t!]
\centering
\includegraphics[width=0.47\textwidth]{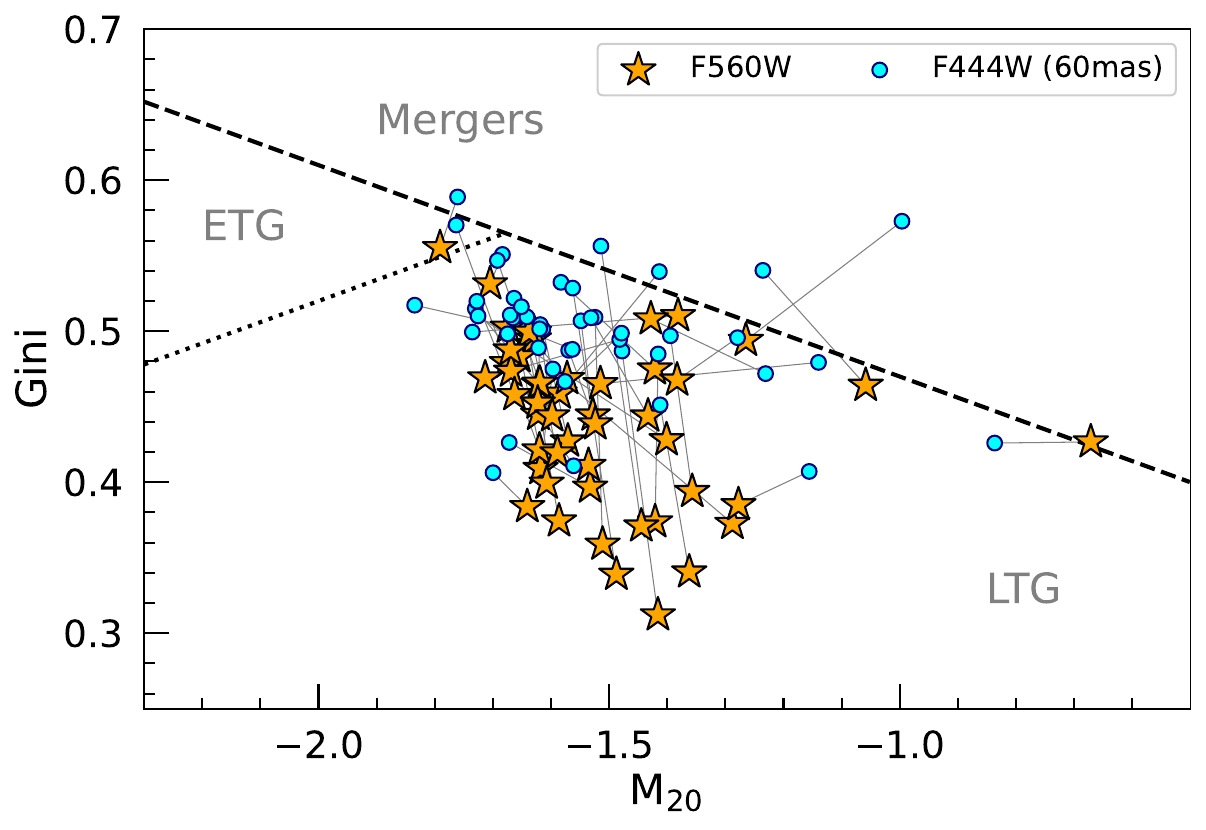}
\caption{As Fig.~\ref{fig:figure2}, but for MIRI F560W (orange stars) and NIRCam F444W images (cyan dots). 
\label{fig:figure_A2}}
\end{figure}

Firstly, we resample NIRCam images to a common pixel scale of 0.06 arcsec
to avoid biases due to the different drizzling of the data \citep[see Appendix~A in][]{Costantin.L:2023a}.
In this case, the PSF $FWHM$ is 0.05 and 0.15~arcsec at 1.5 and 4.4~$\mu$m, respectively.
Then, we measure the non-parametric morphology of each galaxy as
detailed in Sect.~\ref{sec:section3.1} and compare it with the one derived at 5.6~$\mu$m in 
Figs.~\ref{fig:figure_A1} and \ref{fig:figure_A2}.

As expected, we further stress how the structure of galaxies changes
from UV to optical/near-IR wavelengths, transitioning from irregular
to regular morphologies in the Gini-$M_{20}$ diagram.
We derive median Gini$={0.52}^{+0.06}_{-0.04}$ and $M_{20}={-1.3}^{+0.2}_{-0.3}$
at 1.5~$\mu$m, while we measure Gini$={0.50}^{+0.03}_{-0.03}$ 
and $M_{20}={-1.6}^{+0.2}_{-0.1}$ at 4.4~$\mu$m.
The overall distribution of galaxies in Fig.~\ref{fig:figure_A1} is consistent with
the one in Fig.~\ref{fig:figure2}, despite the NIRCam dataset being $\sim4$ times
more resolved. This mainly translates into higher Gini (more concentrated galaxies).
At 4.4~$\mu$m, galaxies show again very compatible $M_{20}$ but
larger Gini coefficient with respect to F560W values
(1\% and 12\% relative difference, respectively),
resulting in slightly more concentrated light distributions,
possibly explained by the combination of different resolutions 
and rest-frame wavelengths probed by NIRCam and MIRI.

\end{appendix}

\end{document}